# Complexity in the Context of Systems Approach to Project Management

Alexei Botchkarev, Ph.D., SMIEE, *Adjunct Professor, Ryerson University*, Toronto, Canada

Patrick Finnigan, P.Eng., SMIEE, *Instruments & Information*, Toronto, Canada.

## Abstract

Complexity is an inherent attribute of any project. The purpose of defining and documenting complexity is to have an early warning tool allowing a project team to focus on certain areas and aspects of the project in order to prevent and alleviate future risks and issues caused by this complexity.

The main contribution of this paper is to present a systematic view of complexity in project management by identifying its key attributes and classifying complexity by these attributes. A "complexity taxonomy", based on a survey of the existing complexity literature, is developed and discussed including the product, project, and external environment dimensions.

We show how complexity types are described through simple real life examples and business cases. Then we develop a framework (tool) for applying the notion of complexity as an early warning tool for a project manager in order to timely foresee future risks and problems.

The paper is intended for researchers in complexity, project management, information systems, technology solutions and business management, and also for information specialists, project managers, program managers, financial staff and technology directors.

## Introduction

For decades, complexity has been acknowledged as a critical project dimension (Baccarini 1996). Since the admitted failure of many large systems, including IT Systems (Charette 2005; Standish 2013; Daniels and LaMarsh 2007) the causes have been widely studied, for example at the mega-project level, where losses are astronomically high, both in terms of cost overrun, but also in the failure of these systems to deliver their critically needed products and strategic objectives. There have been several well- funded and supported efforts to analyze and promote new methods of Complex Project Management (US National Academy of Sciences Transportation Research Board (Shane, Strong & Gransberg 2012); International Center for Complex Project Management Task Force (ICCPM 2013)).

-------------------------------------------------------------------------------------------

These efforts are also driven by increased focus on overall project success, not just the "Iron Triangle" of Function, Cost and Schedule (PMI 2013). It seems that these wider definitions of project success (Howsawi et. al. 2014) also may lead to new methods which can be applied to ensure all aspects of a successful project.

This paper will narrow the focus to a discussion of IT projects/systems and their implementation using some new paradigms for project management, and the overall definition of the success of such complex projects.

In our view, Complexity for IT projects should be defined in a way that will help in focusing on potential challenges stemming from certain types of complexity: thus leading to resolving these challenges. This approach manages projects to get results, not just to classify them for organizational, research, and financial measurement purposes.

"The intrinsic complexity of projects, in part, is driven by political, social, technological and environmental issues, as well as including end user expectations which may change dramatically over the project life- cycle. Indeed, even minor projects can be complicated by hierarchical, siloed, and unnecessarily competitive organisational arrangements, wherein communication and trust can break down" (ICCPM 2013, p.14).

First, we survey the literature on "complexity". Then we present our framework for the understanding and analysis of the behavior of such projects. Then, two case studies of projects we have managed show how the proposed framework for complexity can be applied. Finally, we explore how the framework could be extended and calibrated using data from existing and proposed projects. This will lead to better prediction and management of potential problem areas (aggravated by various aspects of complexity highlighted by our framework), using the proposed paradigm for mitigation of resulting risks.

**What is complexity?**

Pigagaite, Silva & Hussein (2013), indicate that there are "at least 31 definitions of complexity", but the term is often used "because of the lack of a more appropriate expression describing the interrelated features which affect a project's life cycle" - in other words as a catch-phrase for many aspects of systems which we do not understand, or are able to manage.

Intuitively, synonyms for "complexity" are: "intricate", or "difficult". Antonyms include: "simple", "well-understood" and "straightforward". The term has different interpretations in different domains of knowledge such as: computational complexity, systems, biology (Bar-Yam 1997). In the business literature it is often "a state between order and chaos" (Kurtz and Snowden 2003).

Several examples from our everyday experience help us to better understand complexity, and what it is not. In other words, we might describe something as "of large scale", or "containing inherent risks" which are not necessarily complex things. One aspect of complexity is our lack of understanding of the physical laws at work. For example, most users of LED flashlights etc. do not understand the physics and materials science required to manufacture an LED or why in fact an LED can generate such a bright mono- or multi-chromatic (white) light. Because we don't understand the underlying physics, it is easy to label such an LED as "complex". An isolated indigenous individual might have difficulty understanding why an automobile can move with such ease with no apparent human or animal involved – whereas a bicycle is much easier to immediately grasp how a human is involved in making it move. They would describe the automobile as "complex", and perhaps the bicycle as "complicated" etc. but not "complex". In fact, most users of automobiles today would also describe them as "complex", for while understanding the basic mechanics of the engine, transmission etc., most modern automobiles are completely incomprehensible as to the actual components contained in the engine compartment. The shape, size and nature of the sub-components in the engine compartment are a mystery – and the whole is deemed "complex". Individual components which are sealed from view, in particular are regarded as complex – since we can't use our senses, or imagination to deduce their function. The marketing phrase here is: "No user-serviceable components inside". This reaches its zenith with software driven appliances, where the coded logic is hidden from our view or understanding – hence surely a "complex" artefact.

So we can see that there are different concepts of complexity. The variety of definitions hinge around:
- Our ability to mentally decompose the whole into understood parts.
- Our ability to sensually (see etc.) or by mental analogy, deduce the function of the component parts.
- Our understanding may be masked by various intricacies of scale (usually small or large) or the ability of a component to function chemically, electrically etc. outside of our normal understanding and everyday experience in terms of temperatures involved or physical laws at work.
- Our ability to abstractly understand how the individual components work together to function as a more complex assembly. This is a well-accepted definition of complexity - the unpredictable interaction among component parts.
- Our human physiological and psychological limitations ("The rule of seven" items in short-term memory, as with telephone numbers, for example) to deal with several component parts.
- Our lack of correct intuition of how physical systems with energy storage or memory behave (potentially "chaotic" systems).
- In the domain of IT projects, and IT project management, project managers are being asked to deliver systems of unprecedented scope (functionality, number of users, 24x7 operation), which operate globally, and are maintainable across a rapidly evolving technology environment (e.g. versions of web browsers, databases, etc.), while remaining operational through what are sometimes decades of enhancement.

The traditional approach to IT projects and IT project management is based on size, cost, duration. Size is somewhat controversial there can be many participants -- but no "complexity".

A novel approach gaining popularity is based on understanding of the project team as a temporary knowledge exchange group and social networks (Kurtz and Snowden 2003).

Williamson (2011) proposes separating project complexity and project complication, and demonstrates that both have a negative correlation with project success. It should be noted that reducing a complex project and complex IT system into component parts will lead to a reduced understanding of these systems as whole complex entities (Bar Yan 1997).

## Purpose and Scope of the Study

The purpose of this study is to provide analysis of the notion of complexity in the context of systems approach to project management, and develop a framework that can be used by project management practitioners and researchers of the project management methodology to:
   - Identify complexities
   - Understand the challenges they present
   - Deal with the challenges by implementing solutions that alleviate complexities.

The scope of this study is meant to include considerations which are applicable to both private and public sectors (Treasury Board of Canada 2013; Haupt 2003; U.S. National Academy of Sciences Transportation Research Board (Shane, Strong & Gransberg 2012)), at all levels: federal, provincial/state and municipal.

Most considerations of complexity are generic and applicable in any field. However, the focus here is placed on Information Systems/Solutions. Information systems are understood as integrated complexes which include computers (hardware, software), means of communication, people, and business processes, e.g., Enterprise Resource Planning (ERP), Enterprise Content Management (ECM), Business Intelligence (BI) or Customer Relationship Management (CRM) systems.

## Literature Review

We start our literature review with the most popular document used by project managers and project team members - Project Management Body of Knowledge (PMI 2013). The notion of complexity is mentioned many times in the PMBoK 5th edition (PMI 2013). The term *complexity* is used 21 times. Also the adjective *complex* is used 16 times. These numbers don't include three (3) duplicates used in the main text and later in the appendices and glossary. The term *complexity* is mostly used to characterize a project: project complexity or complexity of a project. Often, the term is used in conjunction with project size (the term *size* is also not defined). Most often, *complex* is used in relation to a project (six times), but also it is used as an adjective with products, services, results, processes, procurements.

The PMBoK indicates that several project characteristics depend on complexity (among other attributes):

   - The number of phases, the need for phases, and the degree of control applied (PMI 2013, p. 41).
   - The project management plan's content (PMI 2013, p. 74).
   - The size of the project charter (PMI 2013, p. 74).
   - The applied level of change control (PMI 2013, p. 96).
   - The level of detail for work packages (PMI 2013, p. 128).
   - The cost and accuracy of bottom-up cost estimating (PMI 2013, p. 205).

- The need for formal or informal project performance appraisals (PMI 2013, p. 282).
- Number of stakeholders (PMI 2013, p. 394).

Despite frequent use, the term *complexity* is not formally defined in the PMBoK. Only one indicator or attribute of complexity has been identified. In the project communications chapter, it is mentioned that the number of potential communication channels or paths serves as an indicator of the complexity of a project's communications (PMI 2013, p. 292).

The only recommendation provided to deal with complexity or reduce complexity: iterative and incremental life cycles are generally preferred when an organization needs to manage changing objectives and scope (PMI 2013, p. 46).

Based on the analysis of the usage of the term complexity in the PMBoK, it can be concluded that the term is used to imply the scale of the project (although different from the size of the project). Lack of clarity in the PMBoK regarding the nature of complexity and how to deal with it has negative impact on the practitioners.

In the academic literature, two notions are prevailing in describing project complexity. These notions were stated by Baccarini in arguably the first review paper covering research results on project complexity from late 1960s to mid-1990s (Baccarini 1996). The first notion stems from the systems theory that project complexity can be defined as consisting of many varied interrelated parts. The second notion indicates that difficulty (complicatedness, intricate) is also acknowledged to be used to characterise complexity. However, this attribute was considered subjective and unreliable and in some later publications was separated from complexity narrowing the scope of the phenomenon (Williamson, 2011). Baccarini's (1996) approach is to explicitly define complexity as the numbers of tasks, levels, inputs, etc. This has a positive side in that it tends to be objective and measurable. However, by dismissing complicatedness, these scale attributes tend to miss certain sides of complexity. Relying only on the numbers creates a risk of focussing only on the Size.

The paper by Leukert et al (2012) provides an example of characterizing complexity largely by numeric attributes, e.g. number of users, number of use cases, number of function points, number of user departments, number of infrastructure products (databases, operating systems), number of infrastructure services, number of infrastructure requirements.

It should be noted that when project management practitioners are asked an open-ended question about the main source of complexity, their answer is "the main challenge is the people" (Pigagaite, Silva & Hussein 2013). That testifies to the fact that practitioners' understanding of complexity goes beyond system-only vision and includes complicatedness.

Finally, Coulon, Barki and Pare (2013) present complexity as resulting from unexpected events.

It must be acknowledged (Horgan 1995; Sussman 2000; Sussman 2007; Foster, Kay and Roe 2001; Kurtz and Snowden 2003) that there have been many philosophical discussions on the nature of Complexity, Chaos, and the knowledge-based aspects ("know-ability", "un-know-ability") aspects of so-called "Complex", "Chaotic", "self-organizing", and "non-linear" systems as well as Risk, and Uncertainty in Biology (Solé and Goodwin 2000; Loughlin 2012), Geology (Complex Systems in the Geosciences 2010), Electronics (Axelsson 2002), and Human organizations (Bar Yan 1996) including Healthcare (Haupt 2003; Atun 2012).

The "complexity" in the literature on "complexity" is very broad and directed at a very wide variety of research agendas. Reading this literature does provide a better understanding of complexity issues before narrowing the focus to IT project complexity, and IT project complexity metrics, and it is a lively and intellectually stimulating area of study.

Table 1 shows complexity attributes from a variety of sources with an emphasis on business (i.e. project management), engineering and IT project complexity, across all the life cycle phases: planning, design, creation, or operation and maintenance.

**Table 1 Complexity Attributes**

| **Complexity Attribute** | **Reference** |
|---|---|
| **Structural (Scale)** | Baccarini 1996; Xia and Lee 2004; Geraldi, Maylor & Williams 2011; Albers 2011; PMBOK 2013; Gregory and Piccinini 2013; Turner and Müller 2006 |
| • Number of users. Function | Leukert et al 2012 |
| • Number of use cases, function points. Function | Leukert et al 2012 |
| • Number of user departments. Function | Leukert et al 2012; Turner and Müller 2006 |
| • Multiplicity of geographical locations at which work is performed | Gregory and Piccinini 2013 |
| • Interfaces. Inter-connections | Leukert et al 2012; Albers 2011 |
| • Number of Data Elements | Leukert et al 2012 |
| • Number of Components | Pigagaite, Silva & Hussein 2013 |
| • Number of infrastructure products (databases, operating systems). Technology | Leukert et al 2012; Albers 2011 |
| • Number of infrastructure services. Technology | Leukert et al 2012 |
| • Number of infrastructure requirements. Technology | Leukert et al 2012 |
| **Technological** | Baccarini 1996; Tatikonda and Rosenthal 2000; Gregory and Piccinini 2013 |
| • Technology Novelty (technological newness) | Kim and Wilemon 2003; Pigagaite, Silva & Hussein 2013 |
| • Interdependency of technologies. Interfaces between various systems/subsystems | Pigagaite, Silva & Hussein 2013 |
| **Organizational** | Baccarini 1996; Gregory and Piccinini 2013 |
| **Project Management** | |
| • Size of the project | Turner and Müller 2006; Müller, Geraldi & Turner 2007 |
| • Leadership Style | Pigagaite, Silva & Hussein 2013 |
| • Task Ambiguity | Pigagaite, Silva & Hussein 2013 |
| • Scope changes | Müller, Geraldi & Turner 2007 |
| • Internal complexity of project elements | Ramasesh and Browning 2014 |
| • Lack of robustness of project elements | Ramasesh and Browning 2014 |
| **Uncertainty** | Geraldi, Maylor & Williams 2011; Pigagaite, Silva & Hussein 2013 |
| • Knowable / Unknowable | Kurtz and Snowdon 2003; Gruhn and Laue 2006 |
| • Goals and methods | Turner and Cochrane 1993; Williams 1999 |
| • Environmental uncertainty | Gul and Khan 2011 |
| • People uncertainty (social interactions, rules of interactions) | Gul and Khan 2011 |
| **Ambiguity** (lack of clarity) | Gregory and Piccinini 2013 |
| **End-Users** | |
| • Willingness to adapt. Ability to contribute | Pigagaite, Silva & Hussein 2013 |
| **Dynamics** | Xia and Lee 2004; Geraldi, Maylor & Williams 2011; Gregory and Piccinini 2013 |

| **Pace** (temporal dimension) | Dvir, Sadeh & Malach-Pines 2006; Geraldi, Maylor & Williams 2011 |
|---|---|
| **Constraints** of the objectives, resources or environment | Dunović, Radujković & Škreb 2014 |
| **Socio-political** | Geraldi, Maylor & Williams 2011 |
| • Stakeholders | Turner and Müller 2006; Maylor, Vidgen & Carver 2008 |
| • Diversity of expectations, needs | Pigagaite, Silva & Hussein 2013 |
| • Behavioural, personalities of team members, complexity of interaction | Geraldi and Adlbrecht 2007; Remington and Pollack 2007; Geraldi, Maylor & Williams 2011 |

## Complexity Framework

Since complexity is an inherent attribute of any project, our intention is to define and document complexity in the way that will facilitate building of an early warning tool allowing project managers and teams to focus on certain areas and aspects of the project in order to prevent and alleviate future risks and issues that are complexity-related.

It has been observed that complex projects should be described and investigated as systems-of-systems (SoS) (Gorod, Sauser & Boardman 2008, Zhu & Mostafavi 2014). As the SoS principles, practices and methodology are still being developed, there are no universally accepted approaches or definition.

Our approach is illustrated in Fig. 1. It shows three interrelated systems. The first one, External Environment involves stakeholders external and internal to the company, enterprise with its mission, goals, and objectives, and end users of the information system. There are two main reasons for identifying External Environment as a separate system in SoS: 1. It's importance for the success of the project; 2. Lack of control from the project team over elements of this system. The second system is the Project or internal environment includes activities undertaken to develop or implement an information system. It involves project team and project processes. Finally, the third system is the Product or information system that's being implemented and all of its components or subsystems such as software, hardware, etc. Depending on the specifics of the project, each of the parts may be further decomposed, e.g. if the project has extensive purchasing activities, acquisition may be viewed as a separate system within the internal environment component. These three interrelated systems are used for grouping/clustering complexity attributes.

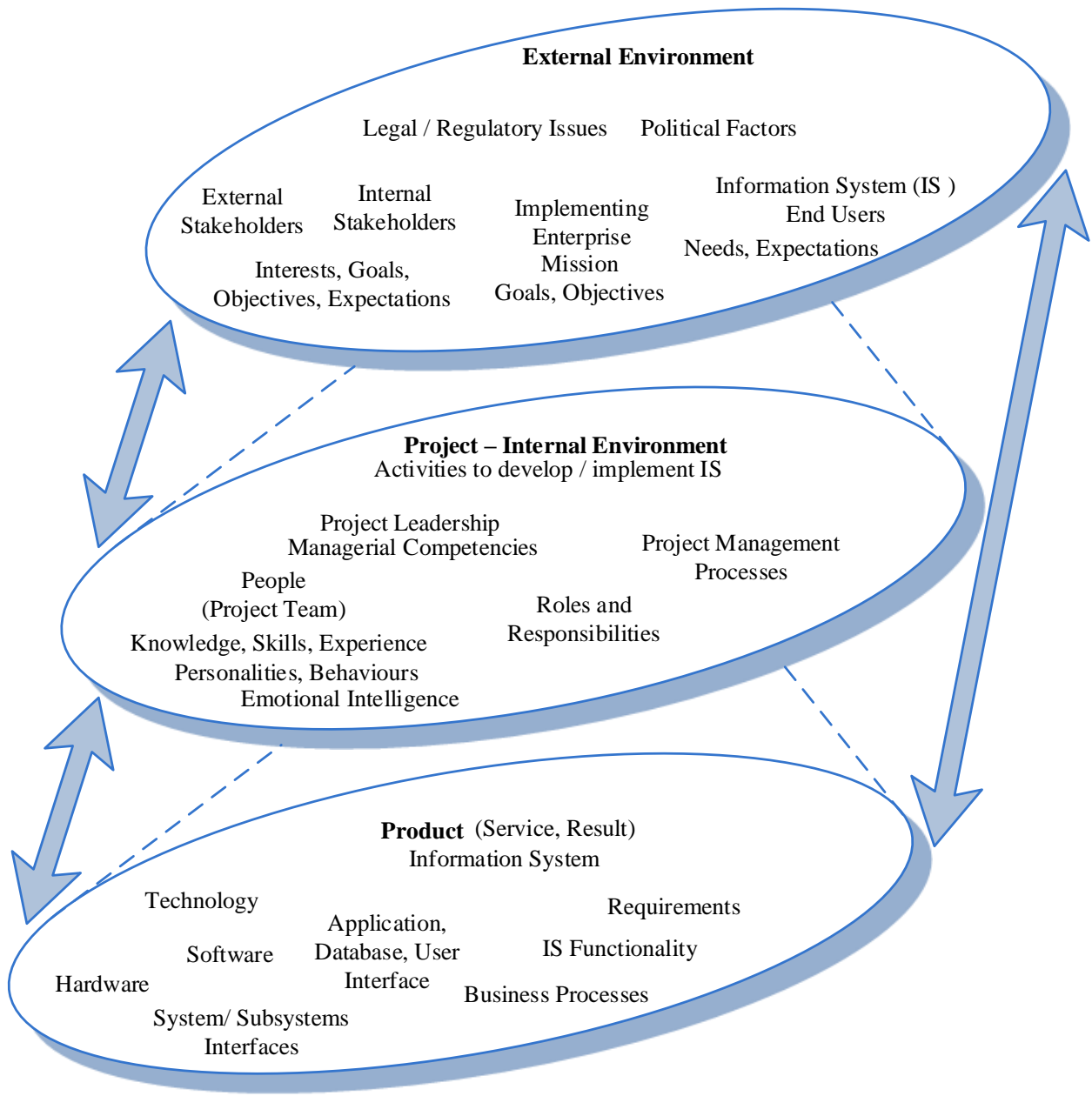

**Fig. 1. Project as a system of systems for complexity mapping**

We propose a paradigm for managing all the specific and interacting aspects of complexity shown in Fig. 2. The paradigm promotes a practitioner-oriented approach. This approach does not only identify the types of complexity, but also reveals the challenges associated with specific complexities and suggests practical steps to alleviate or reduce potential consequences inflicted by these complexities. Complexity attributes are defined in relation to the criteria of project success and project failure factors (e.g. Hussein 2013; Yeo 2002).

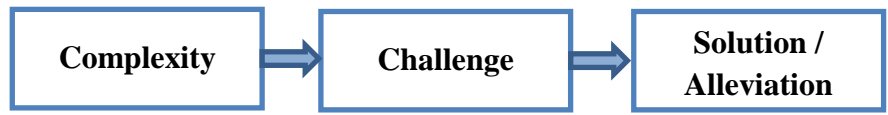

**Figure 2. Paradigm for IS Project Complexity Management**

Literature review revealed that researchers acknowledge the existence of multiple attributes of complexity (summarized in the Table 1). The following criteria were used to select complexity attributes for the purpose of building our framework:
- Systems approach of selecting and grouping attributes.
- Broad understanding of the notion of complexity tending to include complicatedness as well.
- Pragmatic approach of selecting attributes that could be used in project management practice (not only in the academic theoretical constructs).
- Applicability of the attributes to the information systems subject area.
- Attributes commonly present in the IS development and implementation projects.
- Attributes which may have potentially fatal consequences to the project.

Certain complexity attributes (e.g. uncertainty, ambiguity, change, dynamics, risks), commonly referred to as complexities, were not included in the framework (at least at this point) for two reasons. First, these notions have been explored intensively on their own in the academic literature, and there are well-defined tools and practices to deal with them. Second, these notions constitute what could be called "vertical" attributes in relation to our SoS project model – as they pertain to all levels. In this study, we decided to focus on the attributes specific to individual levels of the model for presentation clarity.

The initial version of the completed complexity framework is presented in Table 2. It is self-explanatory, and we'll offer only a couple of comments. For the External Environment, we identified two complexity attributes. Commonly, researchers state that complexity of the project depends on the number of stakeholders and/or end-users (or groups of the above). We argue that the root cause of the stakeholder-related complexities stems from the contradicting expectations/ interests or diversity of goals among stakeholders. We call this complexity attribute stakeholder non-alignment. It is this non-alignment (which may evolve or rise unexpectedly) that the project manager should be carefully monitoring and mitigating. A pure number of stakeholders may not present a challenge, if their interests and expectations are aligned. But even with two or three non-aligned stakeholders on the project things may go sour, if left unattended.

**Table 2 Complexity Framework**

| System of Systems Level | Success criteria | Complexity Attribute | Challenge | Solution/ Alleviation |
|---|---|---|---|---|
| External Environment | • Appreciation by stakeholders<br><br>• Acceptance and appreciation by end-users | **Stakeholder non-alignment.** Weak alignment of stakeholder interests/ goals/ expectations.<br>**User incongruity.** Diversity of user needs and contradictory priorities. Lack of ability to adapt. | Despite the result of the project, there may be no full appreciation. High intensity of these types of complexities may lead to project cancellation. | Early and forthright assessment of interests, expectations and needs. Negotiated, agreed-upon and documented compromises. Continuous monitoring of changes and introduction of adjustments. |
| Project – Internal Environment | • Completion on time<br>• Completion within budget<br>• Complete scope delivered | **Knowledge and skills gaps.** Project team members lack required managerial, technical or project management skills and knowledge. | Planned activities may not be completed and overall scope may not be delivered on time and/or within budget. | Identification/ diagnosing of knowledge and skills shortages early in the project. Targeted knowledge transfer and competencies enhancement. |
| Product (Service, Result) | Achievement of goals/benefits of the final product defined through the performance measures | **Solution (including technology, integration and business processes) challenges.** | Solution may lack required functionality and/or may be under-performing and/or may be non-operational. | Perform end-to-end-testing of the complete solution. |

The framework we present here is not intended to be comprehensive at this point.

This study was focussed on identifying individual attributes of complexity. At the same time, real-life projects are prone not only to a single complexity or several individual attributes of different types, but also to combinations of complexities with unpredictable integral impact

We recommend readers identify and quantify (if possible) each of the complexity attributes shown, and how they interact within your specific projects. Once identified, these complexities can be properly managed.

## Application of the Framework

Two project cases illustrate applicability of the proposed framework. The cases are based on real projects. However, they were rendered anonymous to avoid proprietary and privacy issues.

**Case 1. Project with a combination of complexity attributes**

Context

A large organization was involved in implementation of a CRM system. This Enterprise-wide project included eight business departments, CEO and PMO offices. A total staff of over 400 people were performing three distinct types of operations: information management, investment control and industry liaison.

Complexity

The project had a combination of stakeholder non-alignment and user incongruity complexities.

Challenges

A variety of the interests of diverse business departments led to a "power" struggle between the core stakeholders – departments' heads – regarding the scope and timelines of the project. Diversity of end-user needs complicated the consensus on the initial functionalities.

Solution / Alleviation

Early and forthright assessments of stakeholder interests, expectations and end-user needs were not performed. No stakeholder agreement was negotiated.

Result

This project suffered delays. Executive support was lost. The project has been suspended. Neglected complexities led to suspension of a project which (by the overall agreement) could be very profitable for the organization.

**Case 2: Project with solution challenge**

Context

A large multi-national computer software company has the unique situation of a database product which runs on a number of computer platforms (i.e. large servers, intermediate branch-size servers, and workstations). Customers have come to expect that functionally, it behaves the same across all platforms, although performance is expected to be much faster on the hierarchy of servers. The challenge for this manufacturer is to introduce new versions of the database with significantly enhanced performance to compete against other products, and enhanced functions which will generate new licensing revenue.

The database product is characterized by:
- Scale : 10s of thousands of customers

- History: There are at least 7 previous versions running in production
- Scale: expectation is that the product handles millions of transactions per hour in typical customer applications
- Multiple languages: the product needs to display and store information in most world languages
- Multiple platforms: Quality control testing is required with a variety of new and old hardware platforms and operating systems
- Multi-site development: as is usual in a multi-national software supplier
- Inherent complexity - of a code base of tens of millions of lines of code evolved over many years.

Complexity

The project has a solution challenge combined with complexities of scale.

Challenges

Product management needs to compete for resources with other profitable products in the manufacturer's portfolio. The (largely) environmental complexities of this key component of many customer's larger products needs to be continually emphasized to management, in order that appropriate resources are available for new releases of the product. By showing how critical this product is in many customer environments, customer account executives can see how dependent customers are on this component. By also showing new requirements (for example EU and ISO standards, privacy legislation etc.), it can easily be shown how the "complexity" of a new release of the product is targeted at specific requirements. Resources, risk mitigation etc. can be planned accordingly.

Solution/Alleviation

Project management in this case is best handled as "portfolio management". This is most effective when a management champion is appointed for each of the product versions, whether a new or historical release (in production), or running on a specific platform. Portfolio management is used to prioritize new customer requirements and assess the impact across specific Product, Process and Environment combinations. For example, it may not be necessary to offer all features on all platforms. The various aspects of the product (viz. performance) are also assigned technical champions who plan for and secure resources to ensure their aspect of the product complexity meets the assigned metrics. Assigning the right technical and other leaders to specific aspects of the portfolio complexity ensures the success of that aspect.

A further helpful tool for this product portfolio is the (customer sanctioned) logs of complex transactions against the database. By capturing as much environmental information as possible, as well as the sequence of transactions (read this info, store that info etc.) along with timestamps to unravel the sequence of events, it is possible to diagnose and repair very complex technical issues (i.e. unexpected results or delays).

It is quite viable for a "multi-platform" software product to evolve and thrive even in face of the multiple dimensions of complexity.

## Discussion and Conclusions

By thoroughly documenting potential project complexities within the hierarchy of systems, and fully characterizing them, it may be possible to:

- Secure executive buy-in by better explaining problems arising from complexities, and to various specific complexity mitigation strategies
- Secure funding to implement appropriate risk mitigation strategies based on proven, detailed complexity drivers
- Extend the set of complexity attributes by adding to the framework, based on project experience, and over-arching project mandates such as ISO quality or customer data privacy
- Build questionnaires for staff to more easily characterize and budget for projects (such as (Treasury Board 1, 2, 3), which go beyond and supplement traditional ROI-based project evaluation approaches.
- Develop rules of thumb based on various severities of complexity, and project management heuristics within the context of the project management framework for the management of complexity

A new approach to use the notion of complexity has been proposed - use complexity as an early warning tool for project management within the full context of the framework proposed.

Project complexity is approached from the system of systems methodology.

The contribution of this study includes formation of a new complexity attributes framework based on a revised system of systems layering of projects and proposed complexity-challenge-solution paradigm.

The proposed framework is useful for planning the mitigation of project risks, using the simple paradigm of identifying the complexity, the specific challenge, and documenting and planning for a solution or alleviation of these challenges in order to minimize overall project risk.

The proposed framework has been applied to two project cases, showing practicality of the procedures.

The study has certain limitations. The initially proposed framework contains a limited set of complexity attributes. The conclusions and recommendations of this paper are to be considered in the context of this study, which is that they pertain to complexity in the context of systems approach to project management. The conclusions may or may not be applicable to many other vast and diverse fields where the notion of complexity is used (e.g. engineering, psychology, etc.).

Future research will focus on discovering additional attributes to enrich the complexity framework, and conducting quantitative evaluation of the attributes by involving project management practitioners. Also, future research can be targeted to explore how the framework can be applied to various stages of projects across a wide range of project scale, and for a variety of project methodologies in the context of unique organizational and technical environments. The next edition of the PMBoK should elaborate on complexity and processes to deal with it.

**About the authors:**

Alexei Botchkarev is an Adjunct Professor at Ryerson University in Toronto, Canada. He is a Senior Member of the IEEE. Dr. Botchkarev is a Senior Information Management Advisor with the Health Data Branch, Ministry of Health and Long-Term Care, Ontario. Alexei holds a Bachelor's degree in Electronic Engineering from the Kiev Aviation Engineering Academy (1975), Ukraine, and Ph.D. from the aerospace R&D Institute (1985). Also, he is a Project Management Professional (PMP) certified by Project Management Institute (PMI) and Certified Modelling and Simulation Professional (M&SPCC NTSA/NDIA).

Email: albot@ieee.org

Patrick Finnigan is a Professional Electrical Engineer, and Senior Member of the IEEE. He spent a 40 year career mainly building commercial software products like compilers and databases, but also helping to architect and build several large enterprise-wide software applications. He holds a B.Sc.Physics (York) and an M.Math (Waterloo, 1994).

Email: Patrick_Finnigan@ieee.org